\documentclass[twocolumn]{aastex631}%
\usepackage{enumitem}
\usepackage{hyperref}
\usepackage{booktabs}

\begin{document}
\defcitealias{xu_photometric_2022I}{Paper I}
\defcitealias{xu_photometric_2022III}{Paper III}
\defcitealias{xu_photometric_2023}{Paper IV}
\title{Photometric Objects Around Cosmic Webs (PAC) Delineated in a Spectroscopic Survey. VIII. Revisiting the Lensing is Low Effect}

\author[0000-0002-4574-4551]{Xiaolin Luo}
\affil{Department of Astronomy, School of Physics and Astronomy, Shanghai Jiao Tong University, Shanghai 200240, People's Republic of China}

\author[0000-0002-7697-3306]{Kun Xu}
\affil{Center for Particle Cosmology, Department of Physics and Astronomy,
University of Pennsylvania, Philadelphia, PA 19104, USA}
\affil{Department of Astronomy, School of Physics and Astronomy, Shanghai Jiao Tong University, Shanghai 200240, People's Republic of China}
\affil{Institute for Computational Cosmology, Department of Physics, Durham University, South Road, Durham DH1 3LE, UK}

\author[0000-0002-4534-3125]{Y.P. Jing}
\affil{Department of Astronomy, School of Physics and Astronomy, Shanghai Jiao Tong University, Shanghai 200240, People's Republic of China}
\affil{Tsung-Dao Lee Institute, Shanghai Jiao Tong University, Shanghai 200240, People's Republic of China}
\affil{Key Laboratory for Particle Astrophysics and Cosmology (MOE), and Shanghai Key Laboratory for Particle Physics and Cosmology,  Shanghai 200240, People's Republic of China}

\author{Hongyu Gao}
\affil{Department of Astronomy, School of Physics and Astronomy, Shanghai Jiao Tong University, Shanghai 200240, People's Republic of China}

\author{Hekun Li}
\affil{Shanghai Astronomical Observatory (SHAO), Nandan Road 80, Shanghai 200030, People's Republic of China}

\author{Donghai Zhao}
\affil{College of Physics and Electronic Information Engineering, Guilin University of Technology, and Key Laboratory of Low-dimensional Structural Physics and Application, Education Department of Guangxi Zhuang Autonomous Region, Guilin 541004, People's Republic of China}

\author[0000-0002-8010-6715]{Jiaxin Han}
\affil{Department of Astronomy, School of Physics and Astronomy, Shanghai Jiao Tong University, Shanghai 200240, People's Republic of China}
\affil{Key Laboratory for Particle Astrophysics and Cosmology (MOE), and Shanghai Key Laboratory for Particle Physics and Cosmology,  Shanghai 200240, People's Republic of China}

\author{Chengliang Wei}
\affil{Purple Mountain Observatory, Nanjing 210008, People's Republic of China}

\author{Yu Luo}
\affil{Department of Physics, School of Physics and Electronics, Hunan Normal University, Changsha 410081, People's Republic of China}

\correspondingauthor{Y.P. Jing}
\email{ypjing@sjtu.edu.cn}


\begin{abstract}

The issue of over-predicting the galaxy-galaxy lensing (GGL) signal using conventional galaxy-halo connection models has become well-known as the ``Lensing is Low''  problem, which has been extensively investigated using the Baryon Oscillation Spectroscopic Survey (BOSS) galaxy samples. This issue is also tightly related to the so-called $S_8$ tension.
By applying our Photometric objects Around Cosmic webs (PAC) method to the BOSS survey and the DESI deep photometric survey, we obtained hundreds of cross-correlation measurements to establish an accurate galaxy-halo connection for BOSS galaxies through the halo abundance matching technique (Paper IV).
With this galaxy-halo connection, we show in this work that the predicted GGL signals for BOSS galaxies both in the Planck and WMAP Universes actually agree very well with the GGL measurements. 
We find the best-fitting value $S_8 = 0.8294 \pm 0.0110$, $0.8073 \pm 0.0372$ and $0.8189 \pm 0.0440$ for the CMASS samples with the source galaxies from HSC, DES and KiDS image surveys, respectively. Our work indicates that accurate modeling of the lens population is so critical to interpret the GGL observation.  For the scale of $r_p < 0.6\,h^{-1}\rm{Mpc}$, our GGL prediction for LOWZ samples are also in good agreement with the observations of HSC and DES. However, the GGL observation of KiDS is much lower on the small scale. Our results indicate that no significant baryon feedback is needed to suppress  the small scale clustering unless the the GGL observation of KiDS on the small scale will be confirmed. 
\end{abstract}



\section{Introduction} \label{sec:intro}
Within the theoretical framework of Lambda cold dark matter ($\Lambda$CDM) model, the properties of large-scale structures (LSS) can be predicted with accuracy.
In this framework~\citep{mo2010galaxy,frenk_dark_2012,
somerville_physical_2015,naab_theoretical_2017}, every galaxy forms within a dark matter halo, and the growth of galaxies over time is closely connected to the growth of their host halo.
Therefore, attaining a comprehensive understanding of the statistical correlation between galaxies and halos, commonly referred to as the galaxy-halo connection, is one of the foremost issues of the observed Universe~\citep{wechsler_connection_2018}.

For modeling the galaxy-halo connection, there are two observables, the projected correlation function ($w_p$) and galaxy-galaxy lensing (GGL), both of which are widely used to extract the information in LSS. 
The projected correlation function captures the excess probability of discovering a pair of galaxies as a function of their projected distance, while GGL measures the extent to which foreground galaxies distort the light emitted by background galaxies in their vicinity~\citep{tyson_galaxy_1984,miralda-escude_correlation_1991,brainerd_weak_1996,hudson_galaxy-galaxy_1998}.
The projected correlation function is a traditional tool to constrain the models of galaxy-halo connection and cosmology.
When it comes to the joint analysis of these two observables~\citep{cacciato_cosmological_2013,leauthaud_lensing_2017,abbott_dark_2018,lange_new_2019,singh_cosmological_2020}, the predicted $S_8 = \sigma_8\sqrt{\Omega_m/0.3}$ favor a lower value, approximately 15\% less than the value proposed by \cite{planck_collaboration_planck_2020}, where $\sigma_8$ is the strength of matter fluctuations and $\Omega_m$ is the fraction of the matter-energy density of the matter in the Universe.

Assuming the best-fitting cosmological parameters derived from \cite{planck_collaboration_planck_2020}, meanwhile only fitting the projected correlation function, \cite{leauthaud_lensing_2017} first detected the discrepancy between prediction and observation in GGL.
This study utilized the Halo Occupation Distribution (HOD,~\citealt{jing_spatial_1998,benson_nature_2000,peacock_halo_2000,berlind_halo_2003,zheng_theoretical_2005,zheng_galaxy_2007,contreras_how_2013,guo_velocity_2015}) and Subhalo Abundance Matching (SHAM,~\citealt{vale_non-parametric_2006,conroy_modeling_2006}) models to evaluate the CMASS galaxy samples obtained from the Baryon Oscillation Spectroscopic Survey (BOSS,~\citealt{eisenstein_sdss-iii_2011,dawson_baryon_2013}) and revealed that the observed GGL signal was approximately 20-40\% lower than the predicted value on nonlinear scale (approximately $r_p<1\,h^{-1}\rm{Mpc}$).
This effect, currently known as the ``Lensing is Low'' problem, was also later confirmed by \cite{lange_new_2019} for BOSS LOWZ galaxy samples. 

For the researches about the degree and scale-dependence of this problem, there are large discrepancies between recent studies~\citep{wibking_cosmology_2020,singh_cosmological_2020,yuan_can_2020,yuan_evidence_2021,lange_halo-mass_2021,yuan_stringent_2022,yuan_abacushod_2022} which were conducted with similar HOD model and observational data.
However, \cite{amon_consistent_2023} presented a statistically-significant work by using a standard HOD model with the lens galaxies of BOSS and source galaxies from the Dark Energy Survey year 3 data release (DES Y3,~\citealt{abbott_dark_2018,secco_dark_2022}), the fourth Kilo-Degree Survey data release (KiDS-1000,~\citealt{asgari_kids-1000_2021}), and the Subaru Hyper Suprime-Cam survey year 1 data release (HSC Y1,~\citealt{hikage_cosmology_2019}).
This work found a 20-30\% small-scale discrepancy between predicted and observed signals whose extent showed a trend of reduction towards larger scales.

Several strategies have been adopted without setting a lower $S_8$ Universe, but none succeeded in resolving this issue. 
One such strategy involved galaxy assembly bias~\citep{lange_new_2019,yuan_can_2020,yuan_evidence_2021,yuan_stringent_2022,yuan_abacushod_2022}, which considered additional dependencies on halo properties beyond mass to evaluate the clustering amplitudes of dark matter halos.
Via this strategy,~\cite{yuan_abacushod_2022} found that the ``Lensing is Low'' problem can be alleviated but cannot be completely solved.
When accounting for baryonic effects, the outcomes derived from predictions of hydrodynamical simulations~\citep{leauthaud_lensing_2017,lange_new_2019} or constrained from observations of the thermal and kinematic Sunyaev–Zeldovich effect~\citep{amodeo_atacama_2021,amon_consistent_2023} are insufficient to explain the discrepancy. 
Recently,~\cite{chaves-montero_galaxy_2023} selected IllustrisTNG~\citep{pillepich_simulating_2018} galaxies as mock catalog to mimick the properties of BOSS and found a similar effect as ``Lensing is Low'' problem.
Using a HOD model constrained by the mock catalog, the authors revealed that this discrepancy can be explained by combining the effects from assembly bias, segregation of satellite galaxies relative to dark matter, and baryonic effects on the matter distribution.
In addition, a recent work~\citep{chen_not_2024} find that the ``Lensing is Low'' do not exist in Bright Galaxy Survey and Luminous Red Galaxies targeted by the Dark Energy Spectroscopic Instrument (DESI)~\citep{dey_overview_2019,hahn_desi_2023}.
It suggests that we need more sophisticated and accurate models to describe the galaxy-halo connection for BOSS galaxies.

For BOSS galaxy samples,~\citet[hereafter \citetalias{xu_photometric_2023}]{xu_photometric_2023} presented an accurate measurement of the stellar-halo mass relations (SHMR) and stellar mass completeness based on SHAM with a new method named Photometric objects Around Cosmic webs (PAC).
This method proposed by~\citet[hereafter \citetalias{xu_photometric_2022I}]{xu_photometric_2022I}  builds upon the foundation laid by \cite{wang_galaxy_2011}, enabling the comprehensive utilization of the spectroscopic and deeper photometric surveys by measuring the excess surface density $\bar{n}_2 w_p$ of photometric objects with certain physical properties around spectroscopically identified sources.
Given the significance of small-scale information in addressing the ``Lensing is Low'' problem, we aim to take advantage of the accurate galaxy-halo connection from \citetalias{xu_photometric_2023} to revisit this issue in the paper.

We introduce the methodology of PAC measurement and outline the observational data employed in PAC and GGL measurements in Section~\ref{sec:obs}.
In Section~\ref{sec:modeling}, we describe the methodology utilized to generate mock data in simulations. 
Additionally, we outline the procedure for performing GGL calculation on the mock data.
We present the results in Section~\ref{sec:res} and discuss in Section~\ref{sec:dis}.
We finalize by presenting our conclusions in Section~\ref{sec:con}.

\section{Observational data}
\label{sec:obs}
In this section, we present the observational data utilized in the PAC procedure, and describe the GGL observational data we used.

\subsection{Galaxy-Galaxy Lensing}

We adopt the GGL measurements around BOSS galaxies presented in \cite{leauthaud_lensing_2022}\footnote{\href{https://github.com/alexieleauthaud/lensingwithoutborders}{https://github.com/alexieleauthaud/lensingwithoutborders}}, which compared the measurements of six independent lensing surveys and tested the systematic error.
They adopted the spectroscopic galaxy samples from SDSS-III BOSS DR12 as the lens galaxies.
All the samples were divided into four redshift bins: $0.15-0.31$ and $0.31-0.43$ for LOWZ samples, $0.43-0.54$ and $0.54-0.70$ for CMASS samples, respectively.
For the source catalogs and methodologies in GGL measurements, they adopted six distinct lensing surveys which contained SDSS~\citep{gunn_sloan_1998}, CS82~\citep{leauthaud_lensing_2017}, CFHTLenS~\citep{heymans_cfhtlens_2012}, DES, HSC, and KiDS.

After a blind comparison of the amplitude of $\Delta\Sigma$ (see Equation \ref{eq:delta_sigma}) between the catalogs mentioned above, \cite{leauthaud_lensing_2022} found good agreement between empirically estimated and reported systematic errors which agree to better than $2.3\sigma$ in four lens bins and three radial ranges.
They suggested that the ``lensing is low'' effect can not be fully explained by lensing systematics alone at $z < 0.54$, and thought that the amplitude of  $\Delta\Sigma$ may additionally correlate with foreground stellar density at $z > 0.54$.

In our analysis, we first adopt the GGL measurements of DES Y3, HSC Y1 and KiDS-1000 presented by~\cite{leauthaud_lensing_2022}, and discuss with the corresponding updated data~\citep{amon_consistent_2023}.
\subsection{Photometric objects Around Cosmic webs (PAC)}

The PAC method can estimate the excess surface density $\bar{n}_2w_p$ of photometric objects with certain physical properties around spectroscopically identified sources,
where $w_p(r_p)$ is the projected cross-correlation function (PCCF) and the quantities $\bar{n}_2$ denotes the mean number density of photometric galaxies.

In this study, we directly adopt the measurements of PAC from~\citetalias{xu_photometric_2023}.
The photometric galaxies used in \citetalias{xu_photometric_2023} to get the PAC measurements are obtained from the DECaLS survey, which is part of the Data Release 9 (DR9) of the DESI Legacy Imaging Survey~\citep{dey_overview_2019}. This survey covers a vast region of approximately 9000 $\rm{deg}^2$, encompassing both the Northern and Southern Galactic caps with a declination limit of ${\rm{decl.}} \leq 32^{\circ}$. The spectroscopic samples are from SDSS-III BOSS DR12 LOWZ and CMASS samples, which correspond to two redshift bins: $0.2<z<0.4$ and $0.5<z<0.7$, respectively. The BOSS samples  are selected with ${\rm{decl.}} \leq 32^{\circ}$ to match the footprint of DECaLS.

Based on the galaxy samples mentioned above,~\citetalias{xu_photometric_2023} presented 42 and 33 $\bar{n}_2w_p(r_p)$ measurements in the LOWZ and CMASS redshift ranges, respectively, by dividing the stellar mass of photometric and spectroscopic galaxies into different bins. 
The stellar mass of spectroscopic galaxy is set down to $10^{11.3}\,\rm{M}_\odot$ to make sure a high stellar mass completeness of the samples, while the photometric stellar mass limits are $10^{9.2}\,\rm{M}_\odot$ and $10^{9.8}\,\rm{M}_\odot$ for LOWZ and CMASS redshift ranges, respectively. \citetalias{xu_photometric_2023} modeled the 42 and 33 $\bar{n}_2w_p(r_p)$ measurements in the two redshift ranges using the SHAM method with a parameterized SHMR in an N-body simulation, and achieved a $1\%$ level constraints on the parameters. Then, combining the SHMR and the stellar mass completeness of LOWZ and CMASS samples,  \citetalias{xu_photometric_2023} provided accurate galaxy-halo connection models for the BOSS samples.

For a more comprehensive understanding of the measurement methodology, we refer the interested reader to~\citetalias{xu_photometric_2022I}. 
Additionally, for a detailed description of the data utilized in the excess surface density measurement and modeling, we recommend consulting~\citetalias{xu_photometric_2023}.

\section{Modelings and Predictions of GGL}
\label{sec:modeling}
In this section, we introduce the methodology employed to generate mock data in two simulations with different cosmologies (Planck and WMAP). 
Additionally, we present the method to calculate the predicted GGL signals from mock data.

\subsection{Subhalo Abundance Matching and Stellar Mass Completeness }
Based on the accurate measurements of $\bar{n}_2w_p(r_p)$, one can produce a tight constrain on the SHMR via Markov Chain Monte Carlo (MCMC) method. The parameterized model of SHMR used in this work is a double power law with a scatter:
\begin{equation}
    M_* = \left[\frac{2k}{(M_{\mathrm{acc}}/M_0)^{-\alpha}+(M_{\mathrm{acc}}/M_0)^{-\beta}}
    \right],
\end{equation}
where $\alpha$ and $\beta$ represent the slopes of the high and low mass ends of the SHMR respectively, subject to the stipulation $\alpha < \beta$. 
$M_{\mathrm{acc}}$ is defined as the viral mass $M_{\mathrm vir}$ of the halo at the time when the galaxy was the last central dominant object.
At a specified value of $M_{\mathrm{acc}}$, the scatter in $\log(M_*)$ is parameterized by a Gaussian distribution characterized by the width $\sigma$.
In the study of ~\citetalias{xu_photometric_2023}, the authors adopt the MCMC sampler emcee~\citep{2013PASP..125..306F} to fit this model in  the parameters space $\{M_0,\alpha,\beta,k,\sigma\}$ based on a N-body simulation named CosmicGrowth. 

The CosmicGrowth simulation~\citep{jing_cosmicgrowth_2019} is a suite of high-resolution N-body simulations performed in varying cosmologies through the use of an adaptive parallel P$^3$M code~\citep{jing_triaxial_2002,2021ApJ...915...75X}. Here we adopt a specific simulation of this suite which runs with the cosmological parameters $\Omega_m$ = 0.268, $\Omega_\Lambda$ = 0.732, and $S_8$ = 0.785~\citep{hinshaw_nine-year_2013}. This simulation contains $3072^3$ particles confined within a box with a length of 600 Mpc$\, h^{-1}$.

The identification of groups is performed via the friends-of-friends algorithm, whereby a linking length of 0.2 times the mean particle separation is employed. The subhalos are defined via the HBT+ code~\citep{han_resolving_2012,han_hbt_2018}, which not only identifies subhalos but also provides their evolution histories in this simulation. Furthermore, we utilize the fitting formula outlined in ~\cite{jiang_fitting_2008} to determine the merger timescale for subhalos that comprise fewer than 20 particles (including orphans) and which may remain unresolved. We exclude those subhalos that have already undergone merging with central subhalos.

To enable a comparison with the LOWZ and CMASS measurements, we adopt the catalogs of two snapshots at redshifts of about 0.28 and 0.57, respectively.

In order to check the tension about $S_8$, we adopt an additional set of high-resolution N-body simulations, namely, the Jiutian suite for a higher $S_8$ value. Specifically, we utilize a simulation from this suite which features an impressive $6144^3$ particles and a box size of 1 $h^{-1}\rm{Gpc} $. This simulation is run with the Planck 2018 cosmology~\citep{planck_collaboration_planck_2020}, featuring a set of parameters including $\Omega_m$ = 0.3111, $\Omega_\Lambda$ = 0.6889, and $S_8$ = 0.825. For groups and halos in this simulation, we follow the same procedure utilized in the CosmicGrowth simulation. 

For the LOWZ and CMASS measurements, we use two snapshots from the simulation catalog, taken at redshifts approximately equal to 0.30 and 0.58, respectively.

Regarding the simulation investigated in previous work, we utilize the best-fit directly from ~\citetalias{xu_photometric_2023}. In the case of the Jiutian simulation, we follow the same procedure as mentioned earlier and present the fitting results in Appendix~\ref{sec:apd}.
Besides, we list the corresponding posterior PDFs of the parameters of the SHMR models in Table~\ref{tab:SHMR}.

\begin{table*}
    \caption{Posterior PDFs of the parameters of the SHMR models for CosmicGrowth and Jiutian simulations.}
    \centering
    \begin{tabular}{cccccccc}
    \toprule

    Cosmology&Simulation & Sample & $\log_{10}(M_0)$ & $\alpha$ & $\beta$ & $\log_{10}(k)$ & $\sigma$\\
    \midrule
    WMAP&CosmicGrowth  & LOWZ & $11.579^{+0.012}_{-0.012}$ & $0.429^{+0.006}_{-0.006}$ & $2.215^{+0.022}_{-0.022}$ & $10.105^{+0.011}_{-0.010}$ & $0.201^{+0.003}_{-0.004}$ \\
    WMAP&CosmicGrowth  & CMASS & $11.624^{+0.010}_{-0.010}$ & $0.466^{+0.008}_{-0.008}$ & $2.513^{+0.034}_{-0.033}$ & $10.133^{+0.010}_{-0.010}$ & $0.192^{+0.004}_{-0.004}$ \\
    Planck&Jiutian & LOWZ & $11.641^{+0.013}_{-0.012}$ &$0.433^{+0.006}_{-0.006}$
            & $2.119^{+0.021}_{-0.021}$ & $10.121^{+0.011}_{-0.011}$ &$0.187^{+0.003}_{-0.003}$\\
    Planck&Jiutian & CMASS & $11.681^{+0.011}_{-0.011}$ &$0.438^{+0.008}_{-0.007}$
            & $2.531^{+0.037}_{-0.035}$ & $10.134^{+0.011}_{-0.011}$ &$0.212^{+0.003}_{-0.003}$\\
    \bottomrule
    \end{tabular}
    \label{tab:SHMR}
\end{table*}

Accurate modeling of the SHMR plays a crucial role in effectively describing the galaxy-halo connection, enabling us to capture valuable information even at small scales.
Nevertheless, it is important to consider the stellar mass completeness of the BOSS samples when studying the GGL signals, as the contributions from different mass bins should be precisely determined\citep{leauthaud_lensing_2017}.
In \citetalias{xu_photometric_2023}, we obtained the stellar mass completeness of the LOWZ and CMASS samples by comparing the galaxy stellar mass functions (GSMFs) of the whole population from \citetalias{xu_photometric_2022III} and the GSMFs of the LOWZ and CMASS samples. The model independent measurements of GSMFs from \citetalias{xu_photometric_2022III} are in good agreement with those from SHAM in \citetalias{xu_photometric_2023}, providing a robust and reliable assessment.
In this paper, we adopt the stellar mass completeness of the LOWZ and CMASS samples displayed in the Table D2 of~\citetalias{xu_photometric_2023}.
%at redshift range $0.2<z<0.4$ and $0.5<z<0.7$, respectively.

\subsection{Calculation of Predicted GGL Signals}
Once the mock data was produced, we calculate the predicted GGL signals by estimating the mean surface mass density contrast profile in comoving coordinates:
\begin{equation}\label{eq:delta_sigma}
    \Delta\Sigma(r_p) \equiv \overline\Sigma(<r_p) - \overline\Sigma(r_p),
\end{equation}
where $\overline\Sigma(r_p)$ is the azimuthally averaged and projected surface mass density at radius $r_p$ and $\overline\Sigma(<r_p)$ is the mean projected surface mass density within radius $r_p$.
Here the $\overline\Sigma(r_p)$ can be calculated by
\begin{equation}
    \overline\Sigma(r_p) = \bar\rho_m \int_{-\infty}^{+\infty} \left[
    1+\xi_{gm}(r_p,r_\pi) 
    \right]\,\mathrm{d} r_\pi,
\end{equation}
and $\overline\Sigma(<r_p)$ represents
\begin{equation}
    \overline\Sigma(<r_p) = \frac{2}{r_p}\int_0^{r_p} r'_p \overline\Sigma(r'_p)\,\mathrm{d} r'_p,
\end{equation}
where $\xi_{gm}$ denotes the cross correlation between galaxies and dark matter.
Following the recipe of \cite{leauthaud_theoretical_2011}, we set the $r_\pi^{max} = 50 \,\mathrm{Mpc}\,h^{-1}$ since the $\xi_{gm}$ falls off rapidly enough beyond that scale.

In order to obtain the GGL signals around our mock data, we calculate the cross correlation function $\xi_{gm}$ between our mock galaxy catalog and the dark matter particles in simulation.
Here we adopt the down-sampling particles by a factor of 1000 and  check that this dilution enables sub-percent measurements down to $\sim 0.1\,\mathrm{Mpc}\,h^{-1}$.

Since ~\cite{leauthaud_lensing_2022} calculated GGL signals in physical units, we convert the above units to physical units by using the relationships~\citep{leauthaud_theoretical_2011}
\begin{equation}
    r_{p,ph} = r_p/(1+z_L),
\end{equation}
\begin{equation}
    \overline\Sigma_{ph} = \overline\Sigma(1+z_L)^2, 
\end{equation}
where $z_L$ denotes the redshift of lens galaxies.

\section{Results}
\label{sec:res}
In this section, we present the predicted GGL signals from our galaxy-halo connection model and compare them with observations. 
In order to delve into the enigma of the "Lensing is Low" predicament, we compare the predicted GGL signals from two cosmology with different $S_8$, as exemplified in the study conducted by \cite{amon_consistent_2023}.

\subsection{The variation of GGL along redshift}
\begin{figure*}[ht!]
\includegraphics[width=0.49\textwidth]{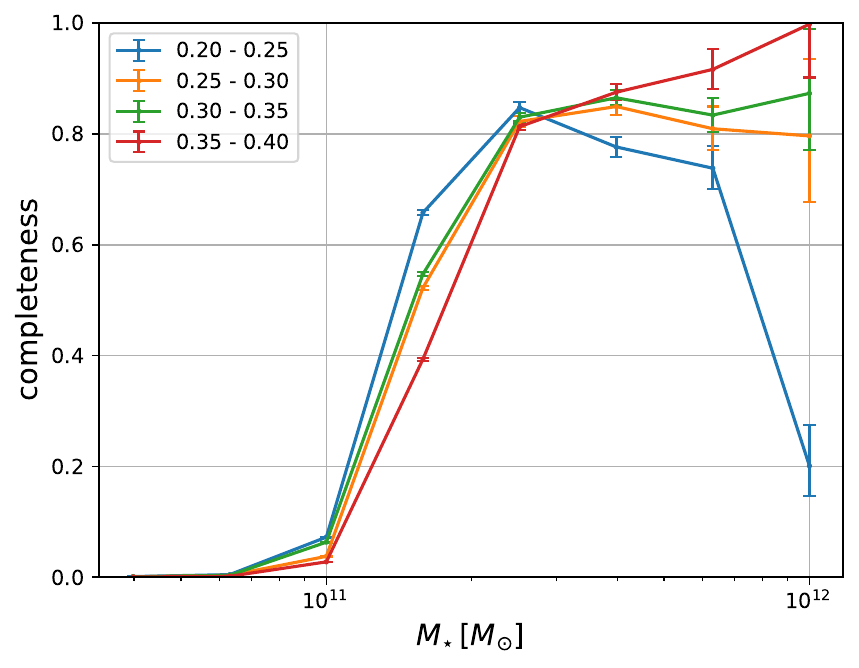}
\includegraphics[width=0.49\textwidth]{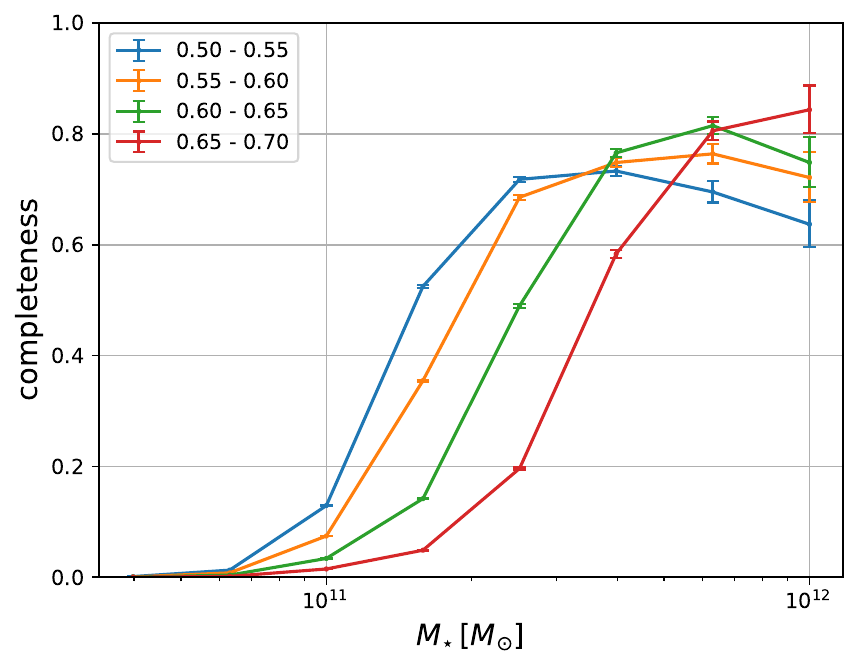}
\caption{Stellar mass completeness of the LOWZ (left) and CMASS (right) samples. The data is provided by Table D2 of~\citetalias{xu_photometric_2023}.
\label{fig:comp}}
\end{figure*}

\begin{figure*}[ht!]
\includegraphics[width=0.49\textwidth]{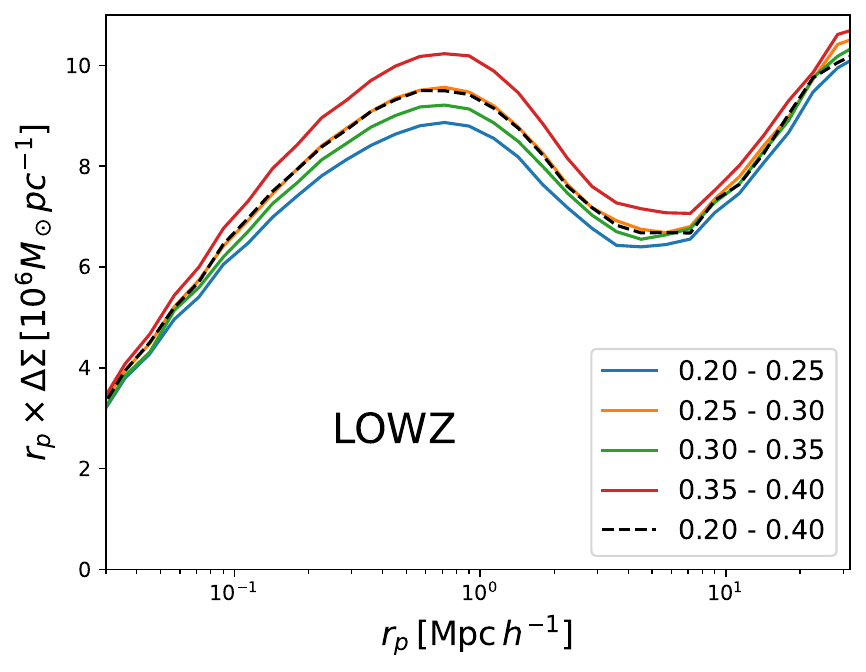}
\includegraphics[width=0.49\textwidth]{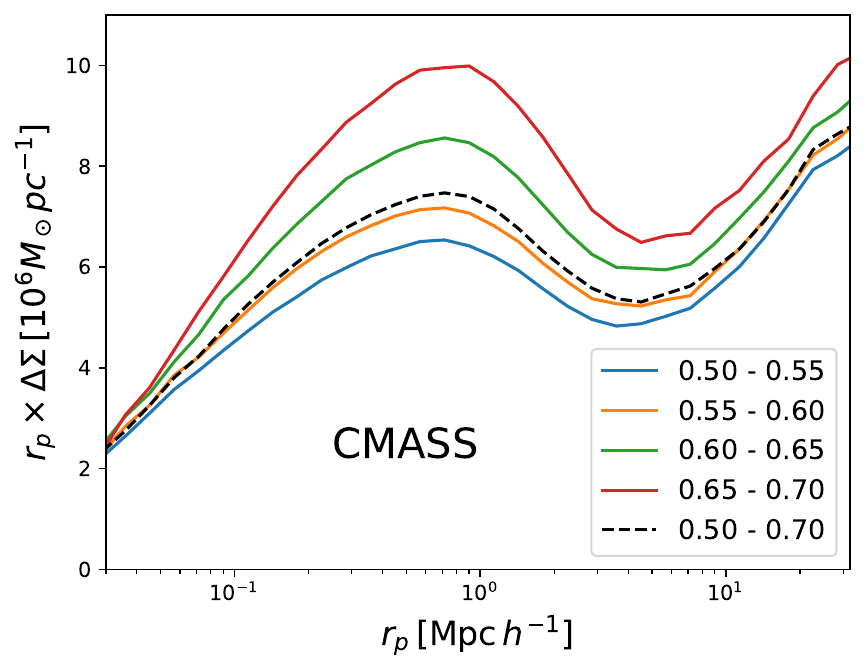}
\caption{The predicted GGL signals in WMAP cosmology, which yield different completeness across redshift bins. The variation of signals along redshift in CMASS (right panel) is much larger than that in LOWZ (left panel).
\label{fig:compz}}
\end{figure*}

As shown in Figure 11 of~\citetalias{xu_photometric_2023} (as same as Figure~\ref{fig:comp} in this paper), the authors found that the stellar mass completeness changes along the redshift.
Moreover, the completeness variations of the CMASS samples exhibit a more pronounced effect across different redshift bins compared to results of the LOWZ samples.
In order to examine the importance of completeness, we display the predicted GGL signals, which yield different completeness across redshift bins in Figure~\ref{fig:compz} with WMAP cosmology.
The dashed black lines denote the signals across the total redshift range $0.20 - 0.40$ and $0.50 - 0.70$ for LOWZ (left panel) and CMASS (right panel), respectively.
We adopt solid colored lines to represent the outcomes of four discrete redshift bins.

In this figure, the variation of the GGL signals along redshift is significant enough to warrant our attention.
In addition, the variation in the LOWZ samples is much lower than that in the CMASS samples, since the completeness varies more significantly in the CMASS samples than in the LOWZ samples.
As a consequence, we suggest that the stellar mass completeness is important in the procedure of generating mock data, especially when we deal with some samples yielding a fast changing of completeness, such as the CMASS galaxies. 

\subsection{Comparison between prediction and observation}
\begin{figure}[ht!]
\includegraphics[width=0.49\textwidth]{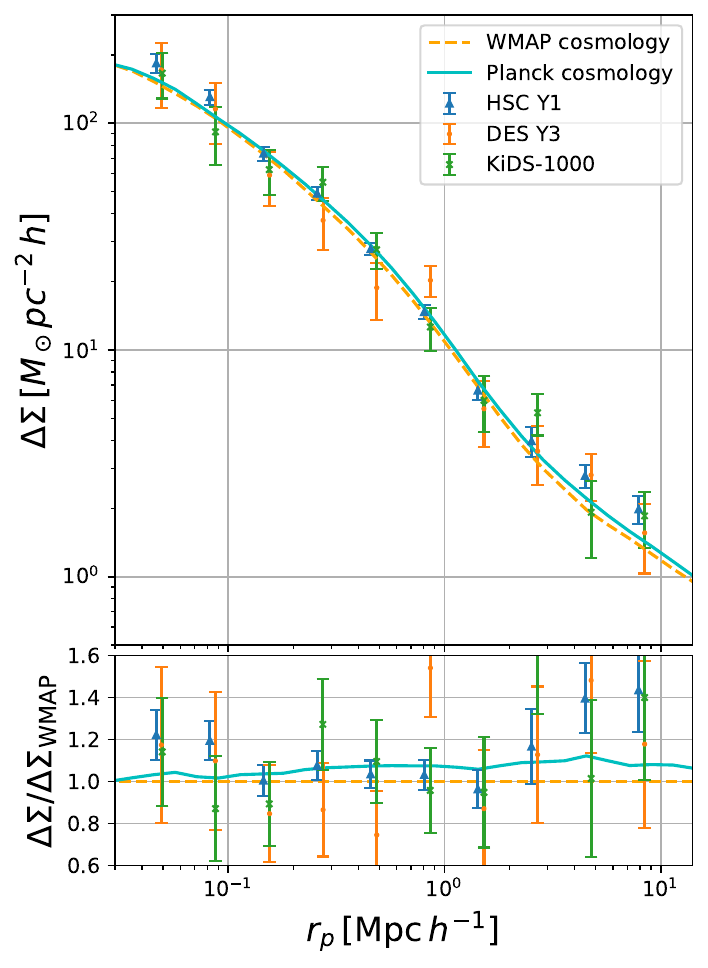}
\caption{Predicted GGL singals in redshift range $0.54 - 0.70$.
The predicted signals are derived from Jiutian (cyan solid line) and CosmicGrowth (orange dashed line) simulations with $S_8 = 0.825$ (Planck) and $0.785$ (WMAP), respectively. The measurements are presented by \cite{leauthaud_lensing_2022} for the CMASS samples with source galaxy from HSC, DES and KiDS surveys in redshift bins $0.54 - 0.70$.
\label{fig:binscompz}}
\end{figure}

\begin{figure*}[ht!]
\includegraphics[width=0.49\textwidth]{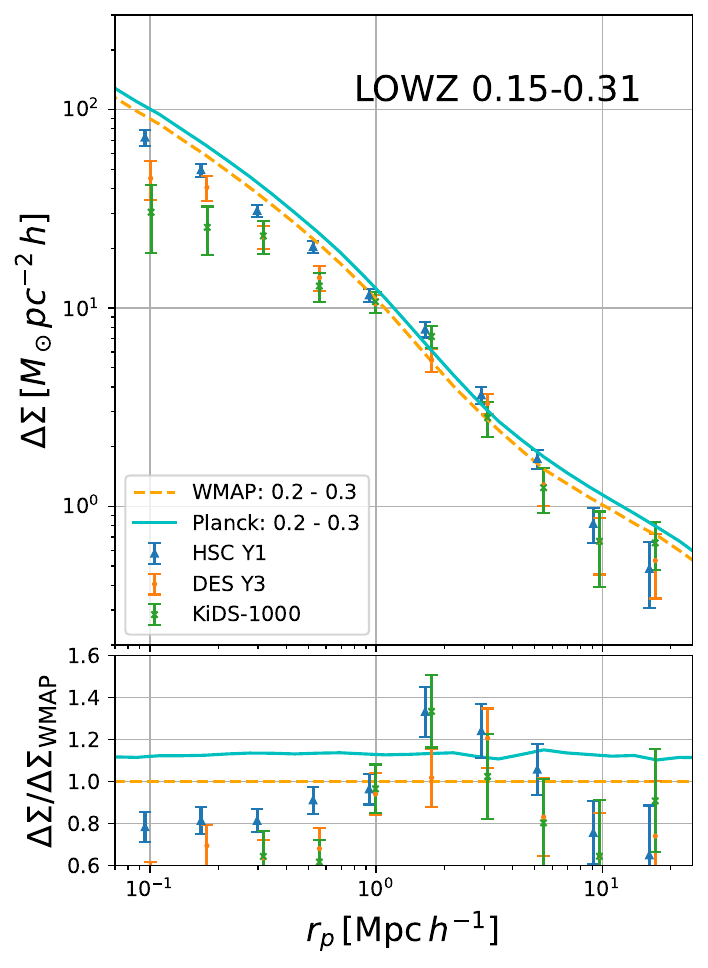}
\includegraphics[width=0.49\textwidth]{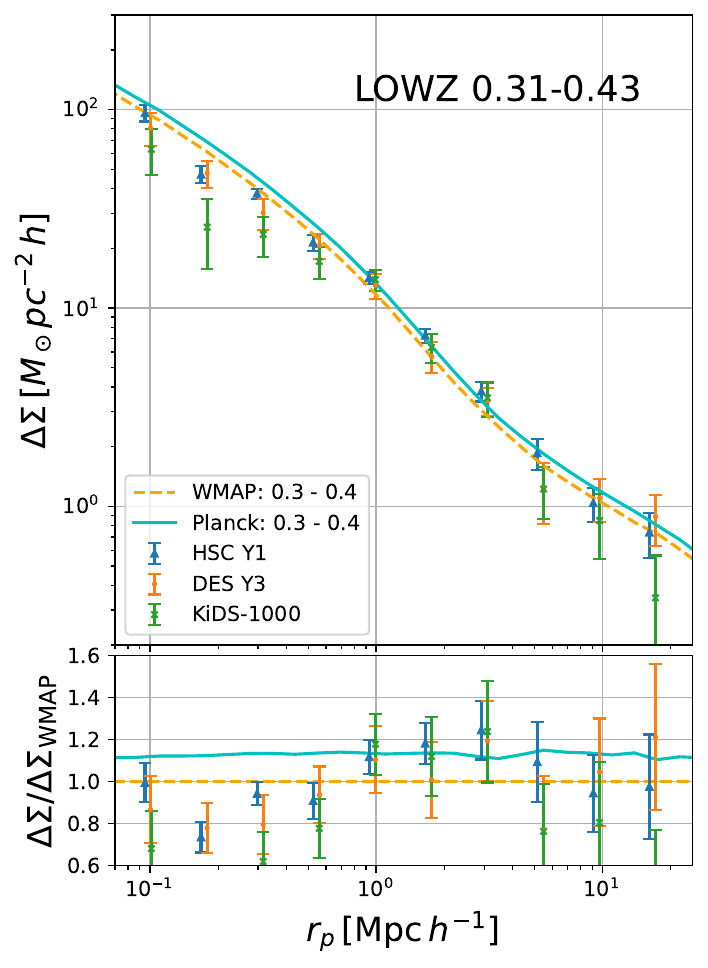}
\caption{GGL measurements for LOWZ galaxies with source galaxy from HSC, DES and KiDS surveys, alongside the predictions of WMAP and Planck cosmology based on the PAC method. 
\label{fig:lowz_hsc}}
\end{figure*}

For the observations of CMASS samples in redshift range $0.54 - 0.70$, we obtain the corresponding incompleteness by interpolating the data in Table D2 of~\citetalias{xu_photometric_2023}.
In Figure~\ref{fig:binscompz}, we display the predicted signals derived from Jiutian (cyan line) and CosmicGrowth (orange line) simulations with Planck ($S_8$=0.825) and WMAP ($S_8$=0.785) cosmology, respectively.
It is obvious that the predictions are not higher than observations across all the scale, the ``Lensing is Low'' problem does not exist in WMAP cosmology or Planck cosmology in this case.

Since the redshift range of the LOWZ samples in our previous work does not contain the observation range, we just consider our predicted GGL signals as an approximation of the effective value in observation for reference.
Figure~\ref{fig:lowz_hsc} shows the GGL results from HSC, DES and KiDS surveys for LOWZ galaxy samples divided into two redshift ranges: $0.15 - 0.31$ (left panel) and $0.31 - 0.43$ (right panel).
We divide the LOWZ samples into redshift range: $0.2 - 0.3$ and $0.3 - 0.4$ to get the predicted signals in this figure.
At $r_p<0.6\,h^{-1}\rm{Mpc}$,  DES Y3 is lower than HSC Y1, and KiDS-1000 is further lower than DES for LOWZ. These discrepancies may indicate some unknown systematics in the GGL measurements at small scales.
In contrast, our predictions derived from the LOWZ samples exhibit a higher degree of consistency at $r_p>0.6\,h^{-1}\rm{Mpc}$.

It is evident that no significant tension exists between observations and predictions.
Both the predicted signals from the Jiutian and CosmicGrowth simulations closely follow the scale dependence of the observed GGL measurements.
Due to the relatively large uncertainties represented by the error bars, it becomes challenging to discern the underlying cosmological distinctions based on the GGL measurements.

Given that the Jiutian simulation employs a higher value of the $S_8$ parameter compared to the CosmicGrowth simulation, the predicted GGL signals obtained from the Jiutian simulation exhibit a commensurate increase in amplitude.
Since the value of $S_8$ holds significant sensitivity to the amplitude of GGL signals, we simply assume that the amplitude of these signals at each scale is linearly correlated to $S_8^2$ and fit the value of $S_8$ with the covariance provided by~\cite{leauthaud_lensing_2022}.

We only fit the $S_8$ value for CMASS samples because the redshift range of our LOW samples do not match the observations.
when considering only the constraints from the HSC survey, we obtain $S_8 = 0.8294\pm0.0110$. 
Conversely, the results of DES and KiDS are quite similar, with values of $0.8073\pm0.0372$ and $0.8189\pm0.0440$, respectively.
All constraints are consistent with our recent work~\citep{xu_accurate_2024} on magnification measurement around CMASS galaxies, which yields the values of $S_8 = 0.816\pm0.024$.
As a result, the combined constraint from all the datasets is found to be $S_8 = 0.8267 \pm 0.0108$. 
It is clear that the largest contribution to the combined constraints comes from the HSC survey, which prefers the Planck Universe.
Considering the constraint from the DES and KiDS survey, the cosmological distinctions cannot be discerned.
In addition, given that we only consider the best-fit value of the SHMR and stellar mass completeness and adopt a simple linear model during the fitting process, the errors mentioned above should be underestimated. 

As a consequence, although the combined constraint is closer to the $S_8$ value of the Planck cosmology, it agrees with both WMAP and Planck within $2\sigma$, making it hard to conclude whether GGL measurements favor a higher $S_8$ Universe with current data. 
Given the variations in their GGL measurements across different weak lensing surveys, caution is warranted when drawing conclusions regarding tensions.

\begin{table}
    \caption{The fitting $S_8$ values for CMASS samples with different datesets.}
    \centering
    \begin{tabular}{cc}
    \toprule
    Survey & $S_8$\\
    \midrule
    HSC  & $0.8294\pm0.0110$ \\
    DES  & $0.8073\pm0.0372$ \\
    KiDS & $0.8189\pm0.0440$ \\
    \midrule
    All  & $0.8267\pm0.0108$ \\
    \bottomrule
    \end{tabular}
    \label{tab:cmass_fit}
\end{table}

\section{Discussion}
\label{sec:dis}
A recent work~\citep{amon_consistent_2023} updates the GGL measurements based on the 
 work of~\cite{leauthaud_lensing_2022}, and they find good agreement between DES Y3, KiDS-1000, and HSC Y1 data.
 Moreover, they combine their measurements from KiDS and DES.
As shown in the Figure 7 in~\cite{amon_consistent_2023}, they suggest that the ``Lensing is Low'' problem can be alleviated by including the baryonic effects.
Their clustering-based HOD predictions for the signal will decrease at small scales by up to $\sim10$ percent and is negligible above 1 $h^{-1}\rm{Mpc}$ with the estimating of the impact of baryons.
This suppression at small scales due to baryonic effects allows their HOD predictions in the context of the Planck cosmology to better match the 1$\sigma$ confidence measurements obtained from HSC.

\begin{figure}[ht!]
\includegraphics[width=0.49\textwidth]{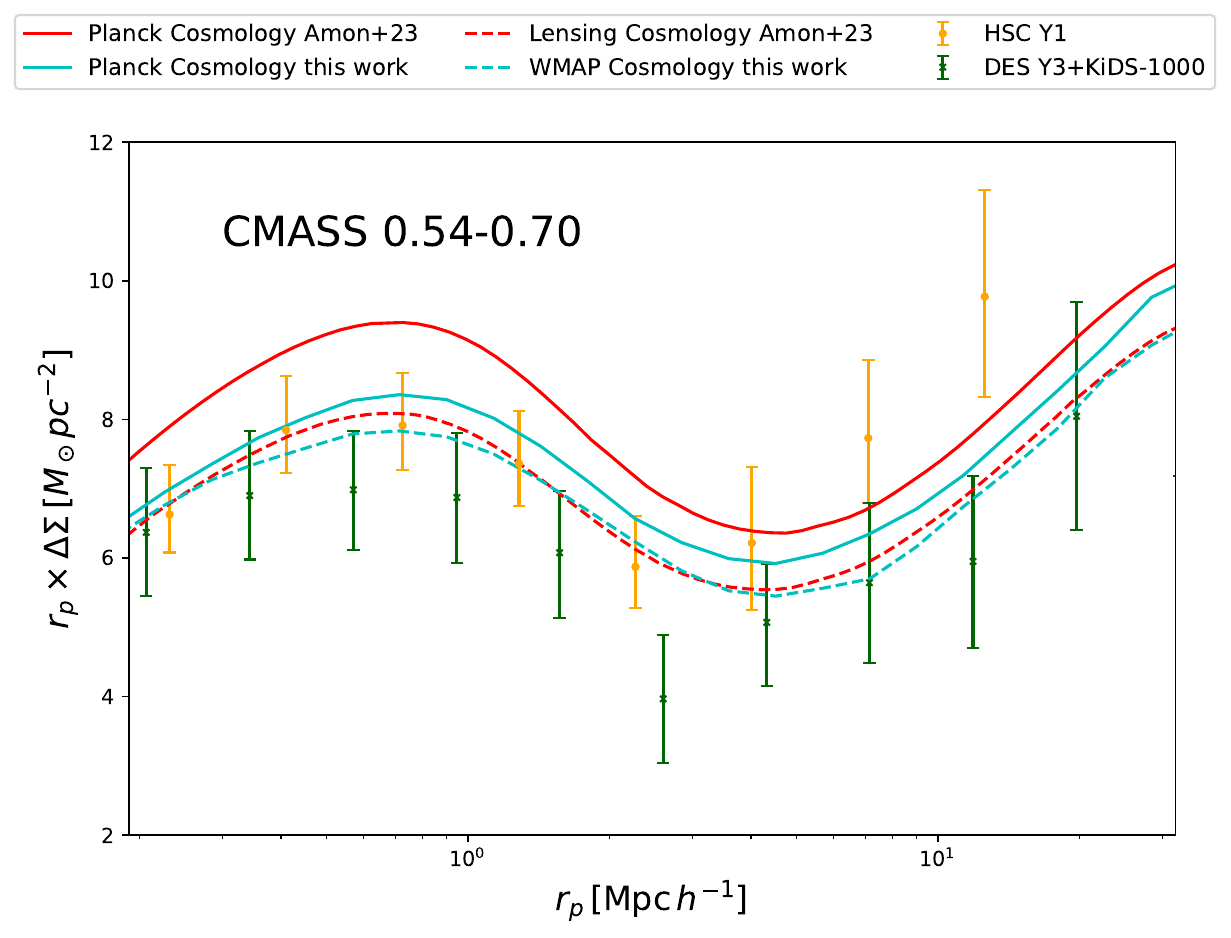}
\caption{Compassion between predicted GGL signals and measured signals using source galaxies from HSC Y1 and
DES Y3 + KiDS-1000 for the results of CMASS samples in~\cite{amon_consistent_2023}.
\label{fig:amoncompare}}
\end{figure}

To provide further elucidation, we present a visual comparison of predicted GGL signals in Figure~\ref{fig:amoncompare}. 
In this figure, we have utilized the same comoving coordinate system to ensure consistency in the representation.
For the CMASS samples, we use red and cyan lines to represent predicted signals of~\cite{amon_consistent_2023} and our work, respectively.
Their results derived from Planck ($S_8 = 0.83$) and Lensing ($S_8 = 0.76$) cosmology are denoted by the solid and dashed lines, respectively.

It is clear that the GGL measurements obtained from DES+KiDS exhibit a slight systematic underestimation when compared to those derived from HSC.
Only their prediction in Lensing cosmology can match the lower measurement in $1\sigma$ confidence level.
As for our works, we find that the GGL signal based on the Planck cosmology ($S_8 = 0.825$) tends to be lower compared to their prediction based on the Planck cosmology ($S_8 = 0.83$). 
This difference is particularly noticeable on small scales, specifically below approximately $1.5\, h^{-1}\rm{Mpc}$.
This decrease in the predicted GGL signal across the entire scale range, as observed in our study, aligns well with the observed data from DES+KiDS, without the need to explicitly consider the impact of baryonic effects.

In summary, our analysis reveals that there is no significant tension observed between the predicted GGL signals and the corresponding observed data.

\begin{figure*}[ht!]
\includegraphics[width=0.49\textwidth]{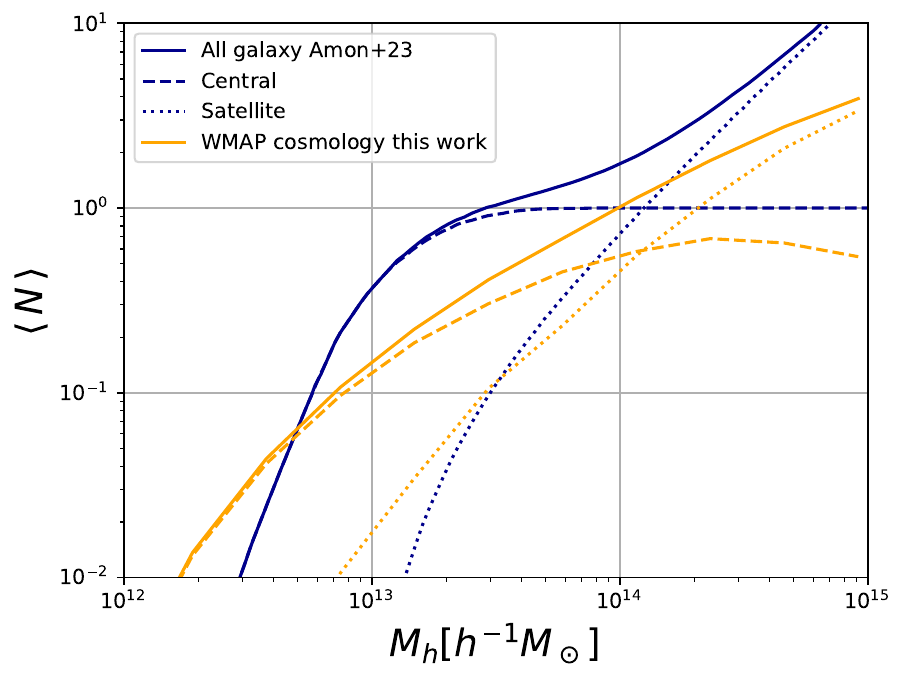}
\includegraphics[width=0.49\textwidth]{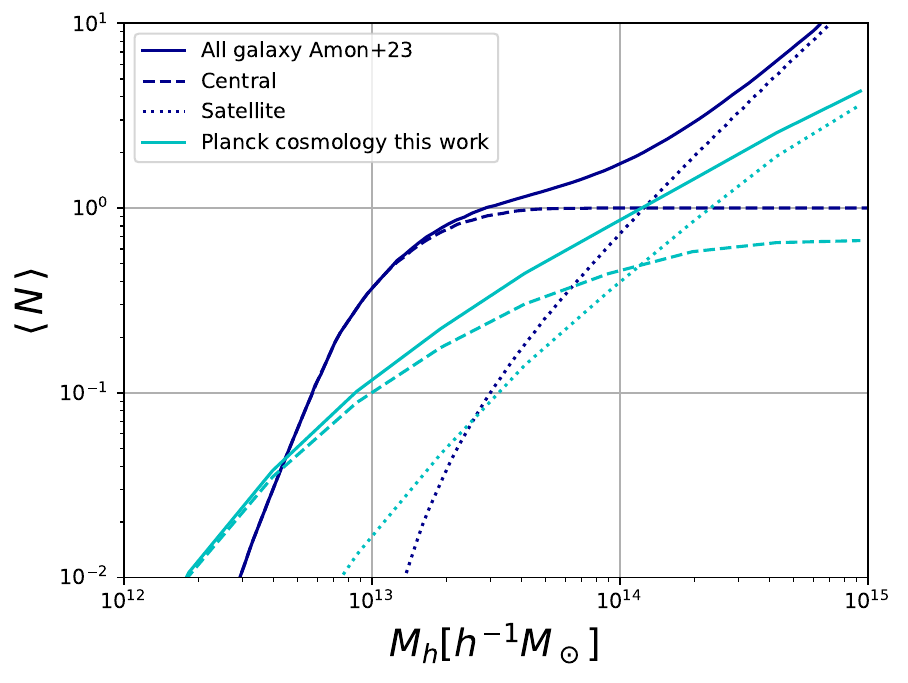}
\caption{HODs for CMASS samples derived from our SHMR after taking the stellar mass completeness into account. The HODs presented by~\cite{amon_consistent_2023} in Lensing cosmology are compared with our results in WMAP (left panel) and Planck (right panel) cosmology.
\label{fig:hods}}
\end{figure*}

To get more comprehensive understanding about the discrepancy in the predicted GGL signals, we compare the HODs of our CMASS mocks with that presented by~\cite{amon_consistent_2023} in Figure~\ref{fig:hods}.
We employ orange and cyan lines to depict the results based the WMAP and Planck cosmology, respectively. The blue lines correspond to the HODs derived from the study conducted by~\cite{amon_consistent_2023} with their Lensing cosmology. Furthermore, the solid, dashed, and dotted lines represent the HODs associated with all galaxies, central galaxies, and satellite galaxies, respectively.

It is evident from the analysis that the HODs derived from the CosmicGrowth and Jiutian simulations exhibit a notable resemblance to each other. However, the HODs presented by~\cite{amon_consistent_2023} demonstrate distinct variations in terms of shape and amplitude.
Comparing to our results, the models in \cite{amon_consistent_2023} prefer more galaxies in large halos.
Due to the higher abundance of central and satellite galaxies within massive host halos, it is evident that the GGL signal experiences an increase on specific scales, particularly in the contribution of the one-halo term on small scales.

Compared with the standard HOD method, our approach combines PAC measurements with the SHAM method in high-resolution N-body simulations. This enables us to accurately constrain the galaxy-halo connection for a given spectroscopic sample, thereby improving our predictions of the GGL signal. 

\section{Conclusion}
\label{sec:con}
In this paper, we explore the ``Lensing is Low'' problem by comparing the GGL measurements of BOSS samples and the predictions using the accurate galaxy-halo connection model based on the PAC measurements and SHAM method in high-resolution N-body simulations.
To generate the corresponding mock data, we employ the SHAM method in N-body simulation via fitting the parameters of SHMR based the PAC measurements of $\bar{n}_2w_p$. 

Since the GGL signal is sensitive to cosmolgical parameter $S_8$, we utilize both the CosmicGrowth simulation with low $S_8 = 0.785$ and Jiutian simulation for a higher $S_8 = 0.825$ to examine this phenomenon. 
For the CosmicGrowth simulation we directly adopt the results of~\citetalias{xu_photometric_2023}, which provides accurate constraint on the parameters of SHMR and stellar mass completeness.
To model the SHMR in the Jiutian simulation, we follow the same process and utilize the same observational data as presented in~\citetalias{xu_photometric_2023}. 
By combining the SHMR and stellar mass completeness, we obtained the galaxy-halo connection for LOWZ and CMASS.

We predict the GGL signals in the context of the WMAP and Planck cosmologies for the CMASS and LOWZ samples based on our galaxy-halo connection and compare them to observations.
Our results can be summarized as follows:

\begin{itemize}
    \item There is no significant tension between the predicted GGL signals from our galaxy-halo connection models of BOSS and the current GGL measurements from HSC Y1, DES Y3 and KiDS-1000. 
    \item The presence of relatively large error in the observations about GGL signal poses a challenge in discerning between the Planck and WMAP cosmologies. Assuming the amplitude of predicted GGL signals are linearly correlated to $S^2_8$, we get the fitting value of $S_8 = 0.8267 \pm 0.0108$ for CMASS samples constrained from the combination of HSC Y1, DES Y3 and KiDS-1000. 
   
    \item  In our study for Planck cosmology, the predicted GGL signals of CMASS align well with the observed data obtained from HSC survey down to 0.2 $h^{-1}{\rm{Mpc}}$. For the lower signals measured from DES+KiDS dataset, our amplitude of GGL signals are slightly higher than that in observation but the shape remains consistent. We suggest that the inclusion of strong baryonic effects in the model may not be necessary.
    \item From the observation side, given the unknown systematics between LOWZ and CMASS, along with variations in their GGL measurements across different weak lensing surveys, caution is warranted when drawing conclusions regarding tensions. 

\end{itemize}

Our findings indicate that the discrepancies observed in earlier research pertaining to GGL are primarily attributable to imprecise modeling of the galaxy-halo connection, or alternatively, to a lack of sufficient flexibility in the models employed, thereby preventing them from fully capturing the clustering information of galaxies on small scale.
Understanding the systematics in the weak lensing surveys is also crucial for solving the ``Lensing is Low'' problem.

\section*{Acknowledgments}
The work is supported by NSFC (12133006, 11890691), by National Key R\&D Program of China (2023YFA1607800, 2023YFA1607801), grant No. CMS-CSST-2021-A03, and by 111 project No. B20019. We gratefully acknowledge the support of the Key Laboratory for Particle Physics, Astrophysics and Cosmology, Ministry of Education. This work made use of the Gravity Supercomputer at the Department of Astronomy, Shanghai Jiao Tong University. 




\bibliography{sample631}{}
\bibliographystyle{aasjournal}

\appendix

\restartappendixnumbering
\section{Fittings of SHMR}\label{sec:apd}

For CosmicGrowth simulation, we directly adopt the SHMR fitting results as provided in~\citetalias{xu_photometric_2023}. These results are visually presented in Figure B6 and B7 of this paper, depicting the SHMR fittings specifically for the LOWZ and CMASS samples, respectively.

\begin{figure}[ht!]
\plotone{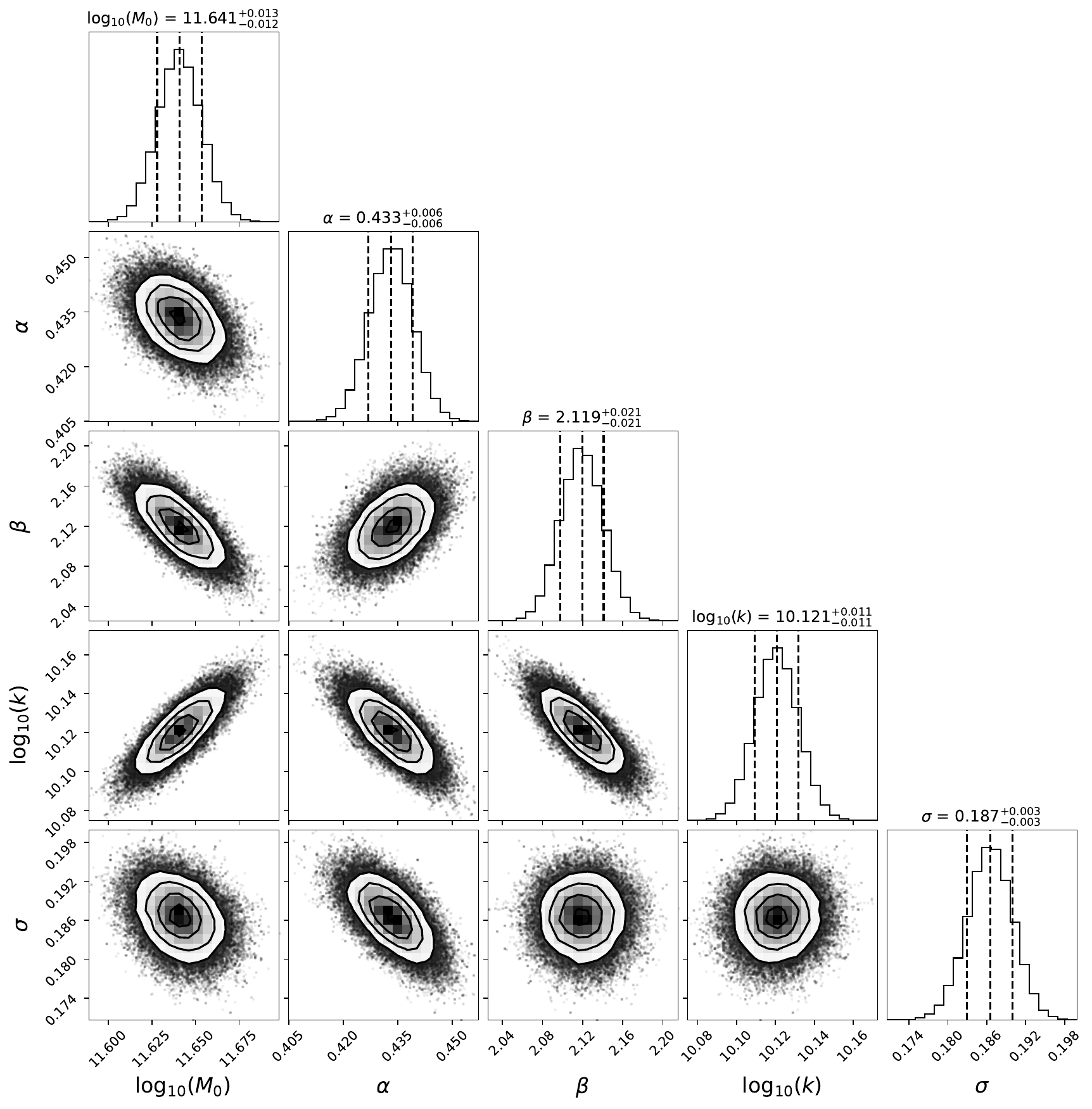}
\caption{Posterior distributions of the parameters of SHMR in Jiutian simultion for LOWZ galaxy samples. The central value is a median, and the error means $16 \sim 84$ percentiles after other parameters are marginalized over.
\label{fig:parmslowz}}
\end{figure}

\begin{figure}[ht!]
\plotone{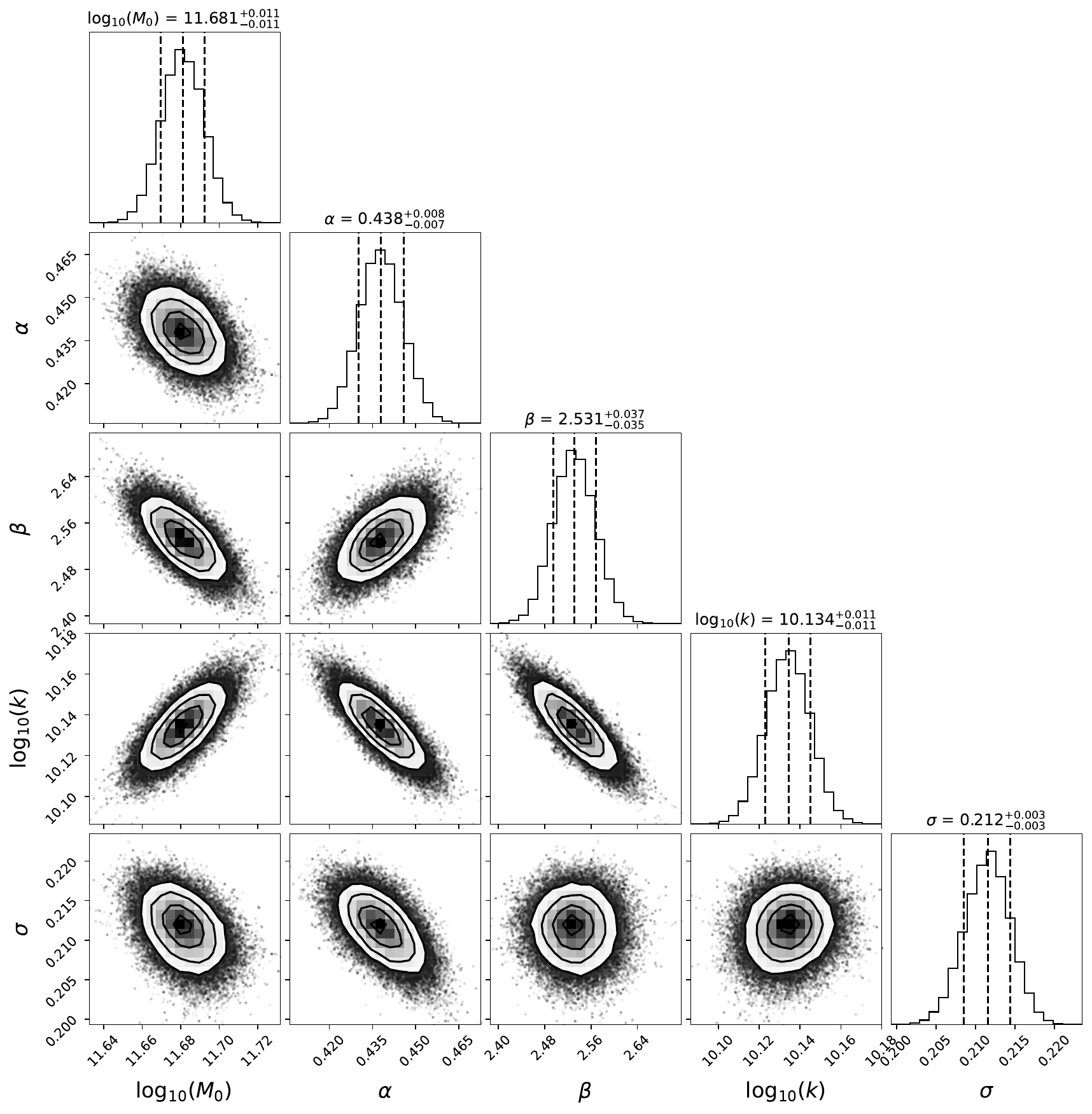}
\caption{Posterior distributions of the parameters of SHMR in Jiutian simulation for CMASS galaxy samples.
\label{fig:parmscmass}}
\end{figure}

In Figure~\ref{fig:parmslowz} and Figure~\ref{fig:parmscmass}, we display the posterior distributions of the parameters of SHMR in Jiutian simulation for LOWZ and CMASS samples, respectively.

\end{document}